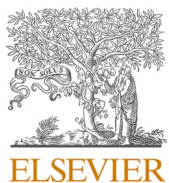
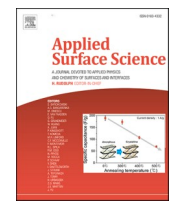

Full Length Article

# Phase purity and surface morphology of high-$J_c$ superconducting Bi$_2$Sr$_2$Ca$_1$Cu$_2$O$_{8+\delta}$ thin films

Sandra Keppert [a,*], Bernd Aichner [b], Rajdeep Adhikari [c], Bogdan Faina [c], Wolfgang Lang [b], Johannes D. Pedarnig [a,*]

[a] *Institute of Applied Physics, Johannes Kepler University, Linz, Austria*
[b] *Faculty of Physics, University of Vienna, Wien, Austria*
[c] *Institute of Semiconductor and Solid State Physics, Johannes Kepler University, Linz, Austria*

A B S T R A C T

Bi$_2$Sr$_2$Ca$_1$Cu$_2$O$_{8+\delta}$ (Bi-2212) thin films with thicknesses less than 50 nm (<20 unit cells) are grown by pulsed-laser deposition (PLD) onto (0 0 1) LaAlO$_3$ (LAO) single crystal substrates. Phase-pure and smooth c-axis oriented Bi-2212 films with optimal oxygen doping, critical temperature $T_{c0}$ up to 86 K, and critical current density $J_c$(60 K) above 1 MA/cm$^2$ are obtained for samples that are annealed *in situ* at temperatures below 700 °C. At higher temperature Bi-2212 films on LAO substrates partially decompose to non-superconducting impurity phases, while films on MgO and SrTiO$_3$ substrates are stable. The broadening $\Delta T_c$ of the metal-to-superconductor resistive phase transition in magnetic fields is much larger for thin films of Bi-2212 as compared to YBa$_2$Cu$_3$O$_7$. The magnetic field-induced suppression of $T_{c0}$ is stronger for Bi-2212 films containing impurity phases as compared to the phase-pure Bi-2212 films. The degradation of LAO substrate crystals after several steps of deposition and chemical removal of the Bi-2212 layer is investigated. New, commercially prepared substrates provide Bi-2212 films with smallest surface roughness (3 nm) and strong out-of-plane texture. However, thin films of almost the same quality are obtained on re-used LAO substrates that are mechanically polished after the chemical etching.

## 1. Introduction

The cuprate high-temperature superconductors (HTS) Bi$_2$Sr$_2$Ca$_{n-1}$Cu$_n$O$_{2(n+2)+\delta}$ (Bi-2212 for $n = 2$, Bi-2223 for $n = 3$) and YBa$_2$Cu$_3$O$_{7-\delta}$ (YBCO) are layered materials with highly anisotropic superconducting and electrical transport properties. The anisotropy can be characterized, for instance, by the parameter $\gamma = \xi_{ab}(0)/\xi_c(0)$ with $\xi_{ab}(0)$ and $\xi_c(0)$ being the Ginzburg-Landau coherence lengths at temperature $T = 0$ K along the *a,b*-lattice plane and the *c*-axis lattice direction of the cuprate unit cells, respectively. Thin film applications exploiting the quasi-2D transport properties of HTS materials require epitaxial layers that are grown on lattice matched substrates or buffer layers. HTS thin films of well-defined unit cell lattice orientation offer several device applications, for instance, as THz radiation emitters [1–4]. Such films are also ideal materials to study the properties of the magnetic flux quanta (vortices) and vortex lattices. Furthermore, artificial pinning centers and defect arrays can be introduced in the films by ion irradiation which enables to tune the superconducting properties of layers [5,6]. In YBCO thin films, different artificial pinning landscapes were created by low-energy light-ion irradiation and effects like unconventional commensurability with maxima of critical current densities $J_c$ for pinning-landscape-specific magnetic field strengths and an ordered Bose glass state of vortices were observed [7–10]. The Bi-2212 thin films are promising to provide more insight into the manipulation and dynamics of vortices due to an even larger anisotropy of Bi-2212 with $\gamma \approx 55$–122 as compared to $\gamma \approx 6$–8 for YBCO [11] and due to the different structure of vortices in the Bi-based cuprates [12–14]. For the creation of nanoscale pinning centers that fully penetrate YBCO thin films with low lateral straggle, the irradiation with He$^+$ ions of energy up to 75 keV has been found to be ideal [15]. Also, for the Bi-2212 thin films He$^+$ irradiation defects with similar structure can be expected provided the film thickness is below 60 nm [8].

For the growth of Bi-2212 thin films, different techniques including pulsed-laser deposition (PLD), molecular beam epitaxy (MBE), sputtering, metal–organic chemical vapor deposition (MOCVD) and sol–gel methods have been employed. A good match of lattice parameters and thermal expansion coefficients of substrate and film and a smooth and stable interface are some of the basic requirements to produce epitaxial Bi-2212 films of high-quality, i.e. with strong in-plane and out-of-plane texture (low mosaicity) and high $T_c$ and $J_c$ values. The growth of high-quality Bi-2212 thin films is more demanding as compared to YBCO films due to the more complex unit cell structure, chemical composition, and phase stability of the Bi-cuprate material. Furthermore, the oxygen doping needs to be precisely controlled to achieve optimum charge carrier concentration in Bi-2212, while overdoping of YBCO is hardly possible. The optimization of Bi-2212 thin film fabrication requires many depositions to be carried out by varying the growth conditions. This, typically, consumes a high number of substrate crystals and therefore the re-use of substrates is an issue. We refer to the procedure substrate crystals undergo before re-use as "recycling". The most important step in recycling is the chemical etching to remove the previously deposited film. This step deteriorates the substrate surface and, therefore, reduces the quality of the thin film that is grown on the re-used substrate [16].

In this paper, we report on the growth and characterization of high-quality Bi-2212 thin films of 25–45 nm thickness by the PLD technique. The parameters of film deposition and post-annealing are optimized to prevent formation of non-superconducting impurity phases. The influence of the substrate on the quality of Bi-2212 thin films is studied, too. Especially, the impact of LaAlO$_3$ substrate surface deterioration after several recycling treatments on the thin film properties is investigated to find a procedure enabling the re-use of such substrate crystals.

## 2. Experimental

The Bi-2212 thin films were grown by PLD as described previously [17]. The substrate temperature was $T = 780\ °C$ and the oxygen background pressure was $p(O_2) = 1.2$ mbar. After deposition the films were annealed *in situ* at $T = 600$ or $770\ °C$ for different time durations at the same oxygen pressure. For comparison, some films were cooled down to room temperature without post-annealing. Single crystals of (0 0 1) MgO with unit cell lattice parameter $a_M = 4.20$ Å, (0 0 1) SrTiO$_3$ (STO) with $a_S = 3.91$ Å, and (0 0 1) LaAlO$_3$ (LAO) with $a_L = 3.82$ Å were used as substrates (MaTecK GmbH). The number of laser pulses applied for target ablation was between 300 and 500, which corresponded to film thicknesses $t_F$ in the range of 25 to 45 nm. The film thickness was measured on lithographically structured tracks by atomic force microscopy (AFM). The laser pulse repetition rate was set to 2 Hz. Bi-2212 ceramics were prepared from commercially available oxide powders by the solid-state reaction method and used as targets for laser ablation. After three depositions, the target surface was sanded to expose pristine and stoichiometric Bi-2212 material to the next PLD process. After 30 depositions the used target was replaced by a freshly prepared ceramic, as the oxygen doping in the target changes due to repeated cycles of heating and cooling in the low-pressure oxygen atmosphere (1.1 – 1.5 mbar). Immediately after deposition, the films were *in situ* annealed at a temperature of either 600 °C or 770 °C and at the same oxygen background pressure as for the deposition. Annealing durations ranged from 30 to 90 min.

The grown films were analyzed with respect to their structural, electrical, and superconducting properties. The surface morphology was investigated via optical microscopy, scanning electron microscopy (SEM), and AFM. The roughness of films ($\Delta t_F$) was the root-mean-square value of the film surface topography as measured by AFM. For determination of the film stoichiometry, energy dispersive x-ray spectroscopy (EDX) and Auger electron spectroscopy (AES) were used. The orientation, phase purity, and mosaicity of the Bi-2212 films was measured by X-ray diffraction (XRD). For the electrical measurements the thin films were patterned into current tracks by UV photolithography and wet-chemical etching. The sample resistance $R(T)$ as function of temperature $T$, the critical current density $J_c(T)$, and the superconducting transition $R(T,B)$ as function of magnetic field $B$ were measured in standard four-probe geometry. The magnetic field was oriented perpendicular to the film surface.

For the recycling of substrate crystals, the previously deposited Bi-2212 films were removed from the LAO substrates by chemical etching in 1 % HNO$_3$ aqueous solution. The substrates were then sonicated in 100 % acetic acid for 3 min and subsequently rinsed with deionized water. Lastly, the crystals were dried by nitrogen gas flush and cleaned once again with acetone. Some LAO substrates were mechanically polished after the chemical etching. Polishing was performed on an automatic specimen preparation unit (Motopol device, Buehler Metaserv Ltd.) in several steps. In the first step a 320-grit emery paper was used and the papers were incrementally changed to finer grits up to 4000. The last polishing steps were done using a diamond suspension solution with granularities of 6 μm, 1 μm, and lastly 0.05 μm. The process finished by polishing with a cotton cloth.

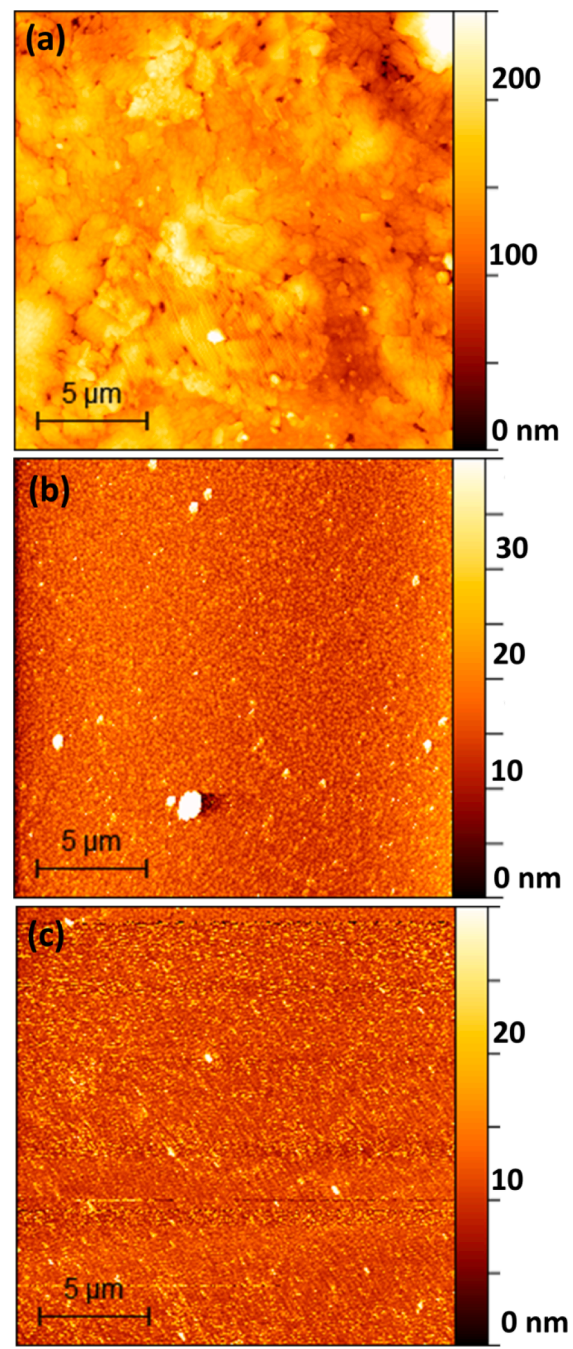

**Fig. 1.** Surface morphology of Bi-2212 thin films on (a) MgO, (b) STO, and (c) LAO single crystal substrates of (0 0 1) orientation (AFM images). Thickness of Bi-2212 films is 100 nm on MgO, 45 nm on STO, and 30 nm on LAO substrates. The film surface roughness is 39 nm on MgO, 4 nm on STO, and 3 nm on LAO.

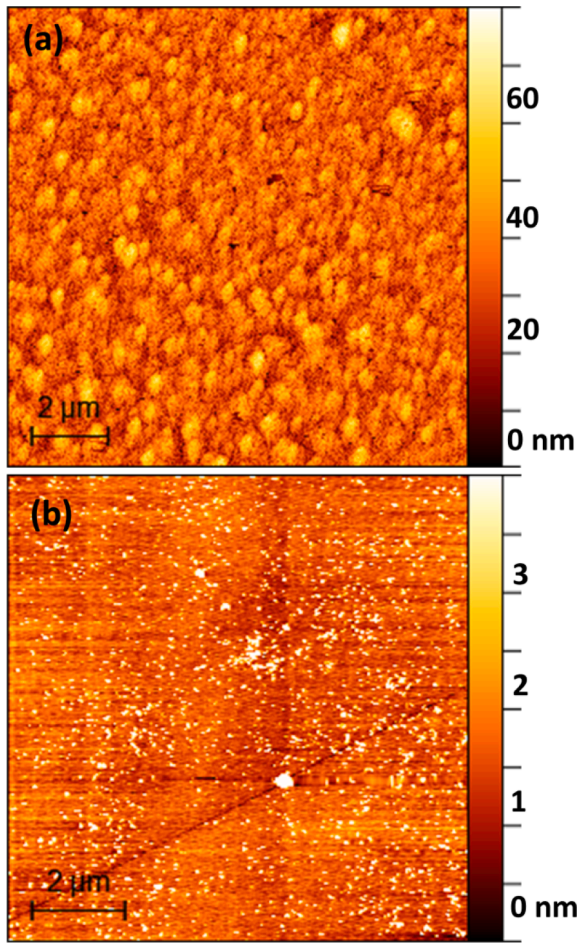

**Fig. 2.** Surface morphology of LAO substrates after (a) chemical etching of a Bi-2212 thin film and (b) chemical etching with subsequent mechanical polishing (AFM images).

## 3. Results and discussion

### 3.1. $Bi_2Sr_2Ca_1Cu_2O_{8+\delta}$ film morphology and crystallinity

The film surface morphology measured by atomic force microscopy (AFM) was found to vary substantially for the different single crystal substrates used (Fig. 1). Bi-2212 thin films of thickness $t_F = 30$ nm deposited on new LAO substrates showed superior surface smoothness and smallest RMS film roughness $\Delta t_F = 3$ nm (Fig. 1c) compared to the other substrates. Films on MgO were thicker as a high number of laser pulses was employed for target ablation ($t_F = 100$ nm). Such films had much higher roughness ($\Delta t_F = 39$ nm, Fig. 1a). Bi-2212 films on MgO with a thickness smaller than 100 nm were inhomogeneous, partially discontinuous and electrically insulating. The discontinuous films were not investigated further as electrical measurements were not possible. The films on STO ($t_F = 45$ nm, $\Delta t_F = 4$ nm, Fig. 1b) had small roughness. The surface roughness correlates with the in-plane lattice mismatch of the Bi-2212 films (unit cell lattice parameters $a = b = 5.40$ Å) and the substrate crystals. For LAO the mismatch $\Delta = |(a - a_L \sqrt{2})/a_L \sqrt{2}| < 0.1$ % is smallest, for STO it is larger ($\Delta \approx 2.4$ %), and for MgO this is very large ($\Delta \approx 9.1$ %) [18].

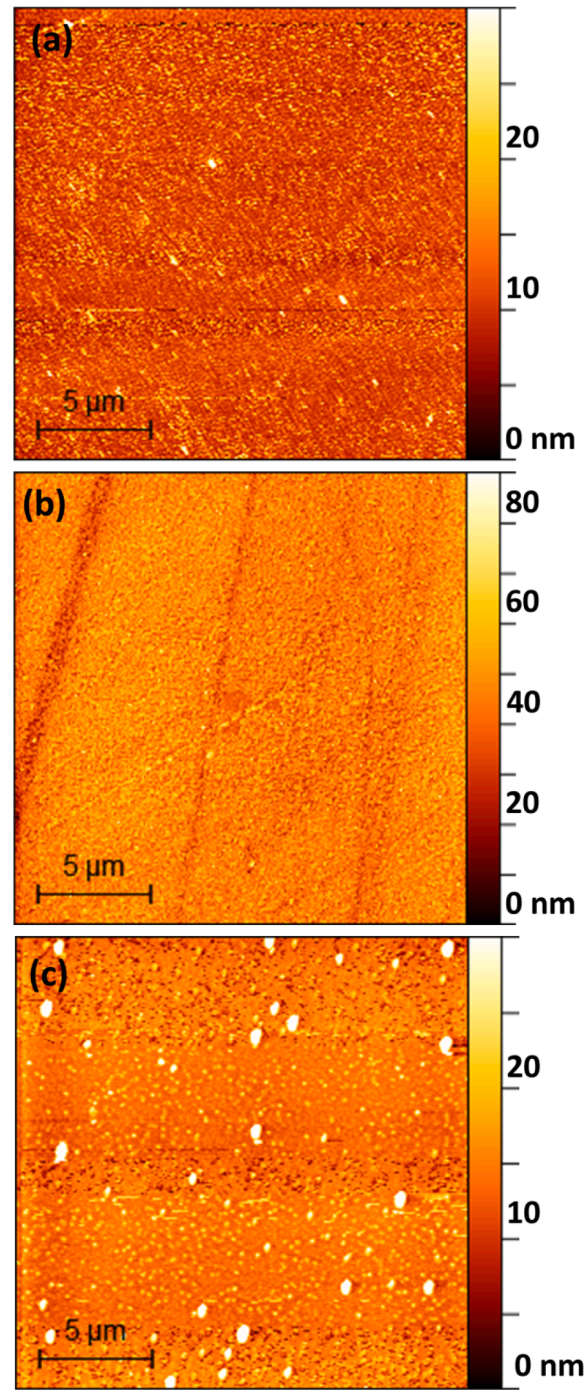

**Fig. 3.** AFM images of 30 nm Bi-2212 films deposited on (a) new, (b) chemically etched, and (c) chemically etched and mechanically polished LAO substrates. The film surface roughness was 3 nm, 7 nm, and 3 nm, respectively.

A dependence of the thin film surface morphology on the treatment of LAO substrates before deposition was observed. Similar effects have been observed with magnetron sputtered YBCO thin films on LAO [16]. For the re-use of LAO substrates, the crystals were repeatedly chemically etched to remove the previously deposited Bi-2212 films. The RMS surface roughness of unused ("new") purchased LAO substrates was<1 nm (data provided by the manufacturer MaTecK). After only one etching treatment, the roughness increased to 9.5 nm (Fig. 2a). SEM/EDX analysis of the etched substrates showed a dense granular surface structure with granule diameter of approx. 10 nm. The EDX results show element specific peaks for La, Al, O and weak peaks from Au (conductive coating for SEM) and a weak peak for C but no sign of Bi, Sr, Ca, and Cu. The etching procedure was sufficient to completely remove the sensitive Bi-cuprate material. The high surface roughness of the etched LAO substrates impedes the growth of smooth layers and very thin films may become electrically insulating as disjointed islands are formed.

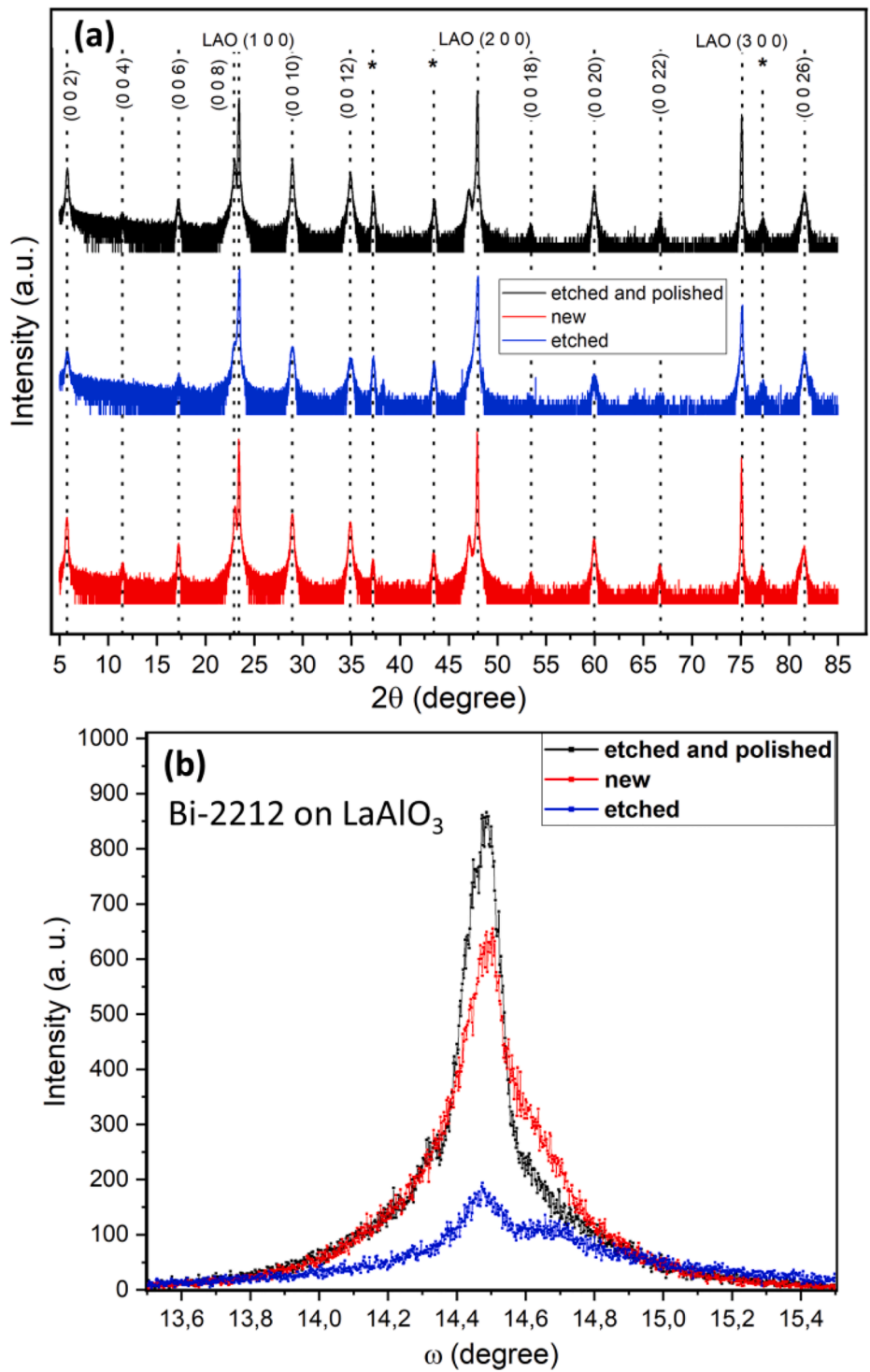

**Fig. 4.** XRD measurements of Bi-2212 thin films deposited on differently treated LAO substrates. (a) Θ-2Θ scans of the *c*-axis oriented and phase pure films on new (red), etched (blue), and etched and polished (black) LAO. Peaks marked with asterisk are from aluminum sample holder. Intensities in logarithmic scale and displayed with offset for clarity. (b) ω-rocking curve of the Bi-2212 (0 0 10) reflection.

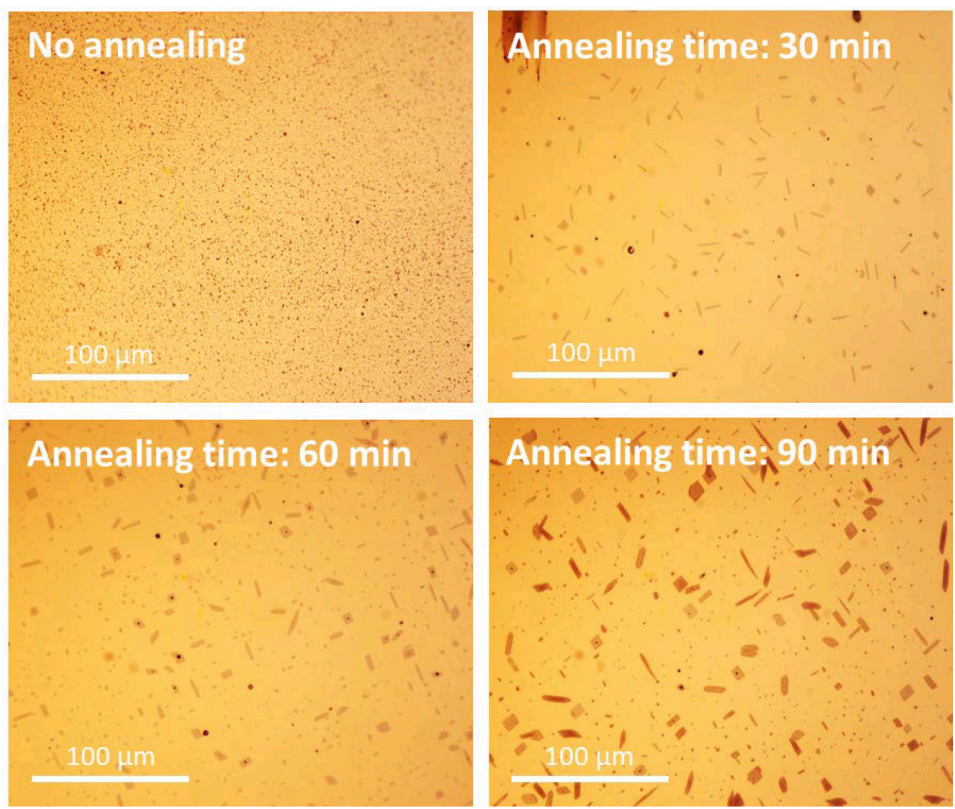

**Fig. 5.** Optical microscopy images (500 times magnification) of Bi-2212 films on LAO *in situ* post annealed at 770 °C for different durations.

Unlike MgO and STO, the LAO substrates undergo a phase transition from rhombohedral to cubic structure at $T = 550$ °C during the deposition process. This leads to crystal twinning and causes surface facets, referred to as "chevron structure" [19,20], which are visible to the naked eye. While unused commercial LAO substrates have a light grey to yellowish color, each heating and cooling cycle crossing the phase transition caused a color change towards a darker brown. This might be caused by accumulation of defects. Also, the substrates were getting more brittle and were markedly more prone to breakage under repeated thermal cycling. Annealing improves the crystallinity of several perovskite substrates by healing defects and relaxing strain [21]. Our LAO substrates were annealed in oxygen flow at 1100 °C for 10 h applying a small temperature gradient (25 °C/h) near the phase transition temperature (550 °C). This treatment showed no improvement regarding the chevron structure and visibly roughened the surface (reduced specular light reflectivity). Therefore, the LAO substrates were mechanically polished after chemical etching which led to better surface quality. The polishing reduced the substrate roughness to 2 nm (Fig. 2b) and eliminated the chevron structure.

The influence of the LAO surface quality on the Bi-2212 thin films was investigated by AFM and XRD measurements. Fig. 3a shows the surface of a 30 nm thin Bi-2212 film deposited on a new LAO substrate resulting in an RMS surface roughness of 3 nm (AFM image). Films deposited on an etched LAO substrate had a surface roughness of $\Delta t_F = 7$ nm (Fig. 3b) while on an etched and polished LAO substrate the Bi-2212 films had a roughness of 3 nm (Fig. 3c). Surface roughness values were calculated excluding particulates on the film surface which were a result of the PLD process and not a result of the substrate surface quality [22]. As low surface roughness of superconducting films has been linked to improved critical current densities, the Bi-2212 films should be as smooth as possible [23].

XRD measurements were performed on Bi-2212 films deposited on new, etched, and etched and polished LAO substrates. The XRD Θ-2Θ scans revealed all films to be *c*-axis oriented with no foreign phases present (Fig. 4a). The intensities are shown with offset for clarity (logarithmic scale). The $(0\,0\,\ell)$ diffraction peaks of the Bi-2212 phase are measured with high signal/noise ($\ell \leq 26$). Several of the diffraction angles of the cuprate films ($\ell = 8$, 16, and 24) and of the LAO substrates ($h = 1$, 2, and 3) were nearly overlapping as the *c*-axis length of Bi-2212 ($c = 30.78$ Å) is closely matching an integer multiple of the lattice parameter of LAO ($c = 8.06\ a_L$). The peak intensities for films grown on new and recycled substrates were about the same and higher than for films on the etched LAO substrates.

For a rigorous analysis of the out-of-plane texture of the thin films, ω-rocking curves of the Bi-2212 (0 0 10) peak were measured. The width of this diffraction peak was $\Delta\omega_{FWHM} = 0.224°$ for films grown on new LAO substrates (Fig. 4b, red curve). Films on etched LAO crystals showed much smaller diffraction intensities, a double-peak structure, and much larger width. However, on substrates that were etched and polished before deposition of a new Bi-2212 film, the width $\Delta\omega_{FWHM} = 0.152°$ was even better compared to films on the pristine LAO (Fig. 4b, black curve).

From these results we conclude that highly *c*-axis textured and smooth Bi-2212 thin films can be grown on new and on properly recycled (i.e., etched and polished) LAO substrates. XRD measurements of Bi-2212 films on STO and MgO substrates (thickness 100 nm) showed very similar phase purity, lattice orientation, and texture [17].

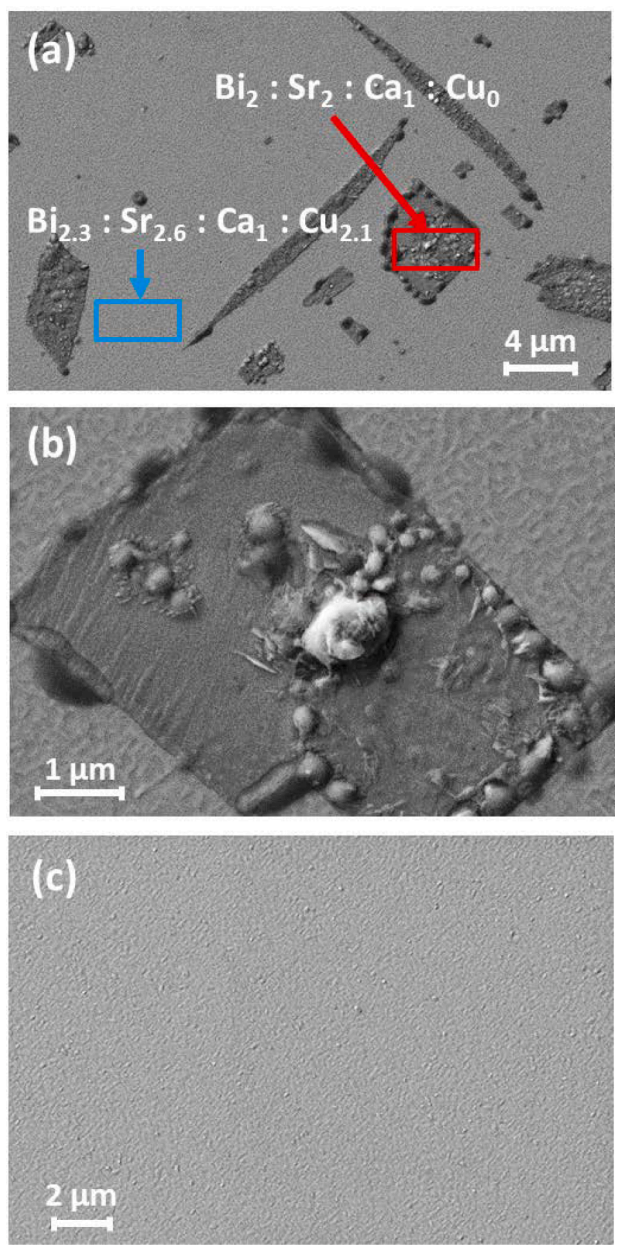

**Fig. 6.** SEM images of Bi-2212 thin films annealed *in situ* after PLD. Impurity structures in films annealed at 770 °C (a, b). Stoichiometry of marked areas (a) measured with EDX. (b) Closeup of an impurity structure. (c) Bi-2212 film annealed at lower temperature showing no impurity.

### 3.2. Phase impurities of Bi-2212 thin films on LaAlO$_3$

Bi-2212 films were deposited at substrate temperature $T = 780$ °C and *in situ* annealed or cooled down to room temperature immediately after deposition. The annealing was necessary to adjust the oxygen doping level δ of films. Fig. 5 shows optical microscopy images of Bi-2212 films on LAO substrates that were post-annealed at $T = 770$ °C for different time durations. Films cooled down right after deposition, i.e. without annealing, showed particulates on the surface as is frequently observed with PLD [22]. Films that underwent annealing showed modified surfaces. The particulates disintegrated and new structures ("impurities") were formed. Such impurities only occurred during the annealing process and grew larger in number and size with annealing time. They were evenly distributed over the whole film area. No impurity was visible for films cooled to room temperature immediately after deposition. Notably, Bi-2212 films on MgO and STO substrates deposited and annealed with the same parameters did not show such impurities. This is in agreement to the phase diagram of bulk Bi-2212 which shows a stable 2212 phase up to about 820 °C at an oxygen pressure of 1 mbar [24,25]. At lower oxygen pressure or higher temperature the 2212 phase becomes unstable and it decomposes into $Bi_2Sr_2CaO_x$ and $CuO_2$ [24,25]. For the thin films of Bi-2212 on LAO, however, we found a lower decomposition temperature at the oxygen pressure of 1.2 mbar.

In order to investigate the impurities in Bi-2212, the films were measured by XRD, SEM/EDX and AES. The XRD patterns did not show any differences to films without impurity. This result points to an amorphous structure or a small volume fraction of the impurity phases, leading to weak signals indistinguishable from the XRD detector noise. SEM and EDX measurements were performed to study the morphology and composition of annealed films (Fig. 6). The impurities have a square or needle like shape with sharp boundaries to the surrounding film matrix (Fig. 6a). EDX analysis revealed that the impurity had the composition $Bi_2Sr_2Ca_1O_x$ while the film matrix showed a stoichiometry close to Bi-2212 (Fig. 6a, b). Due to the lack of copper, no superconducting Cu–O planes are present in the impurities and they are therefore detrimental to the film quality. In fact, this $Bi_2Sr_2Ca_1O_x$ impurity phase has been reported to be insulating [25]. As the sensing depth in EDX is much larger (∼μm) than the film thickness (30 nm) and no other signal was detected the impurity phase most probably extends over the complete layer thickness.

In order to avoid the impurity formation, the annealing temperature of films on LAO has to be lowered to 600 °C. Even short annealing timespans of only 10 min at temperatures above 600 °C led to impurity growth. Impurity-free Bi-2212 thin films on LAO are obtained by annealing at 600 °C for 90 min (Fig. 6c).

AES mapping was performed to determine the spatial distribution of chemical elements in the samples (Fig. 7). The image brightness corresponds to the relative concentration of the respective chemical element. The SEM images taken shortly before (Fig. 7a) and after (Fig. 7g) AES show an impurity of rectangular shape. The AES maps show rather even distribution of Oxygen (Fig. 7b) and Bismuth (Fig. 7c) over the whole sample which is not connected with the shape of the impurity. The concentration of Calcium (Fig. 7d) and Strontium (Fig. 7f) in the impurity is increased while the concentration of Copper is strongly reduced (Fig. 7e) compared to the film matrix.

The relative element concentrations in film matrix and impurity as measured by EDX (Fig. 6) and by AES (Fig. 7) do not agree for Sr and Ca. The EDX-measured concentrations are slightly higher for Sr and the same for Ca in the film matrix. The AES-measured concentrations of both elements are lower in the matrix. This deviation is probably due to a modified Bi-2212 stoichiometry at the film surface and the different sampling depths of the two surface analytical techniques employed. In EDX, the thin film is measured over its total thickness as the escape depth of X-rays is few micrometers, typically. On the other hand, the escape depth of Auger electrons is few nanometers, typically, and only the film surface is contributing to the measured AES signal. However, for Cu and for Bi the EDX and AES results agree confirming the lack of copper in the impurity. After the AES mapping some new particles were found within the impurity structure (SEM images in Fig. 7a and Fig. 7g). These particles were mainly composed of Bismuth (Fig. 7c).

The lack of Cu in the impurity is probably due to diffusion of Cu into the surrounding film matrix. A fast diffusion process could explain the even Cu concentration in the film and the absence of a spatial concentration gradient around the impurity (Fig. 7e). Evaporation of $Cu_2O$ and CuO phases that could have formed is unlikely at low temperature (770 °C) [26]. The amount of Cu-free impurity material is relatively small compared to the total sample volume (Fig. 5). The expected slightly higher concentration of Cu in the film matrix was not detected by the EDX measurement, which has an accuracy of only about 5 at%.

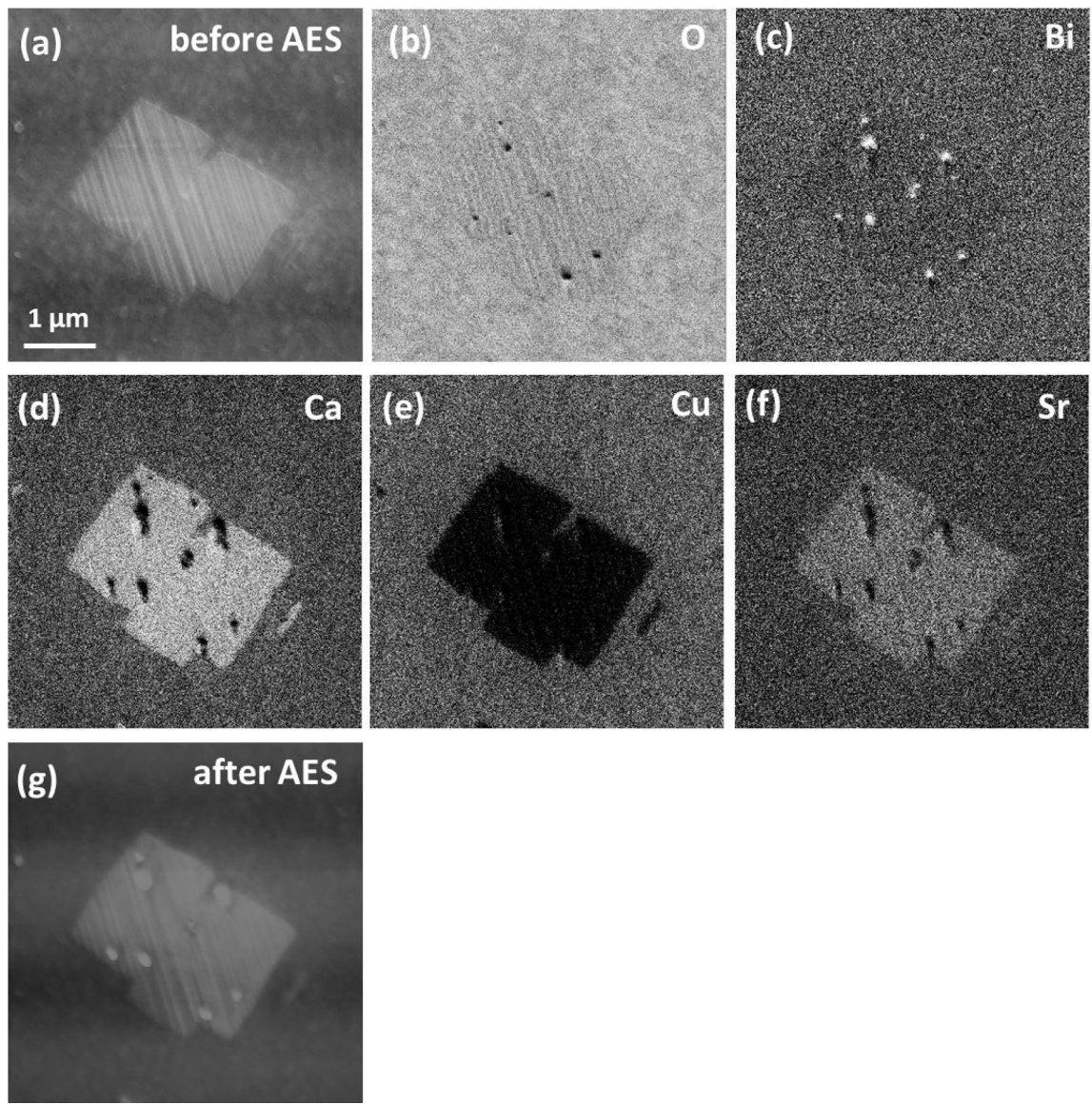

**Fig. 7.** AES elemental maps of Bi-2212 thin film with rectangular impurity structure. SEM image before (a) and after (g) AES measurement. Chemical maps for elements O (b), Bi (c), Ca (d), Cu (e) and Sr (f).

**Table 1**
Surface roughness $\Delta t_F$ and critical temperature $T_{c0}$ of Bi-2212 thin films grown by different techniques.

| HTS | Substrate | Technique | $\Delta t_F$ (nm) | $T_{c0}$ (K) | Ref. |
|---|---|---|---|---|---|
| Bi-2212 | STO | Sol-gel | 10.7 | 81 | [23] |
| Bi-2212 | MgO, STO | MBE | <1 | not disclosed | [27] |
| Bi-2212 | MgO | PLD | < 10 | 78 | [28] |
| Bi-2212 | MgO | Sputtering | 16.2 | 75 | [29] |
| Bi-2212/ 2223 | STO | MOCVD | 1.17 | 89 | [30] |
| Bi-2212 | LAO | PLD | 3 | 86 | this work |

### 3.3. Superconducting properties of Bi-2212 films

The impurity-free Bi-2212 films on new and on recycled LAO substrates that were post-annealed at 600 °C for 90 min in 1.2 mbar oxygen had the best superconducting properties. The highest zero-resistance critical temperature of optimized Bi-2212 films ($t_F$ = 42 nm) was $T_{c0}$ = 86 K. For different samples grown under the same, optimized parameters on such substrates the $T_{c0}$ was in the range from 80 to 86 K. The variation in critical temperature is probably due to the narrow PLD parameter windows (substrate temperature, background pressure) and experimental uncertainties in parameter control. For comparison, the Bi-2212 films grown on LAO substrates that were only etched and not polished had low $T_{c0}$ < 77 K in most cases. Such films of inferior superconducting quality were not further investigated. Films containing impurity structures had lower critical temperature and lower critical current density. A comparison regarding $T_{c0}$ and surface roughness of Bi-2212 films deposited on different substrates by various deposition techniques is given in Table 1 (refs. [23,27–30]).

The self-field critical current density of a phase-pure Bi-2212 thin film and a film with impurity on LAO substrates is shown in Fig. 8. The $T_{c0}$ of the phase-pure, optimized and of the non-optimized films was 83.2 K and 82 K, respectively. Both samples were patterned identically (inset Fig. 9a) and the $J_c$ values were determined using a voltage criterion of $V_c$ = 100 nV for both samples.

The critical current density of the phase-pure film was higher than that of the impure film. For example, at temperature $T = 60$ K the $J_c$ was 1.21 MA/cm$^2$ and 0.83 MA/cm$^2$ for the impurity-free and the impurity-containing sample, respectively. The impurities reduced the cross section area of the superconducting material and thereby the $J_c$ value. The superconducting properties ($T_c$, $J_c$) of Bi-2212 films on STO and MgO substrates (film thickness 100 nm) were reported earlier [17]. Thinner films on MgO ($t_F < 100$ nm) were electrically insulating and films on STO ($t_F = 40$ nm) had a very low critical temperature of $T_{c0} \approx 50$ K.

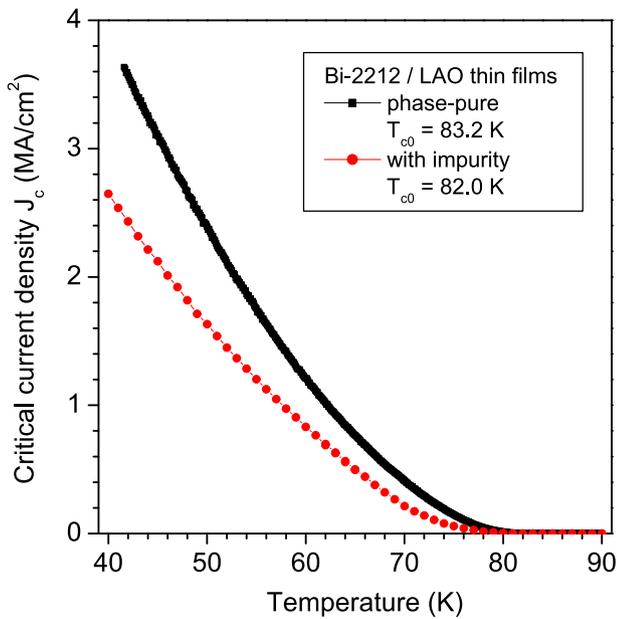

**Fig. 8.** Self-field critical current density of Bi-2212 thin films on LAO substrates. Film thickness was 45 nm for the phase-pure film and 41 nm for the film with impurity.

The electrical resistance versus temperature $R(T,B)$ of samples was measured at different magnetic field strengths (Fig. 9). The films were lithographically structured (inset of Fig. 9a) and the orientation of the applied field $B$ was parallel to the $c$-axis of Bi-2212. The $R(T,B)$ curves for an optimized Bi-2212 thin film are shown in Fig. 9a. The onset temperature of the superconducting phase transition was $T_c^{On} \approx 100$ K and at zero field the critical temperature was $T_{c0} = 83.2$ K. The in-plane resistivity at room temperature was $\rho_{ab} = 0.46$ mΩ cm and close to the value measured on Bi-2212 single crystals ($\rho_{ab} = 0.2$–$0.8$ mΩ cm, depending on oxygen doping level) [31,32]. In magnetic field the width of the transition was increased due to a reduction of $T_{c0}$.

The field broadening of the resistive transition has been observed in YBCO and other cuprate HTS materials. This phenomenon is discussed in terms of thermally assisted flux flow (TAFF) and the activation energy for the decoupling of flux vortices from pinning sites can be derived from fitting TAFF model functions to measured $R(T,B)$ curves [33]. Results from measurements of a patterned Bi-2212 thin film with impurity structures are shown in Fig. 9b. The room temperature resistivity was $\rho_{ab} = 1.1$ mΩ cm. The critical temperature in zero field was $T_{c0} = 82$ K and a similar broadening of the metal-to-superconductor transition was observed as for the optimized film. The onset temperatures of both samples were similar. From the transition broadening the upper critical magnetic field of the Bi-2212 thin films at zero temperature $B_{c2}(T=0)$ was estimated [34]. The phase-pure Bi-2212 film had $B_{c2} = 111.6$ T which was higher than the critical field of the film containing impurities ($B_{c2} = 94.2$ T). The Ginzburg-Landau coherence length as calculated from the $B_{c2}(0)$ fields of films is $\xi_{ab}(0) = 17$–$19$ Å and very close to the reported value of 19 Å [11,35]. These findings imply that the superconducting matrix in the impurity-containing film has similar qualities to the optimal film, but the impurities reduce the amount of the superconducting material.

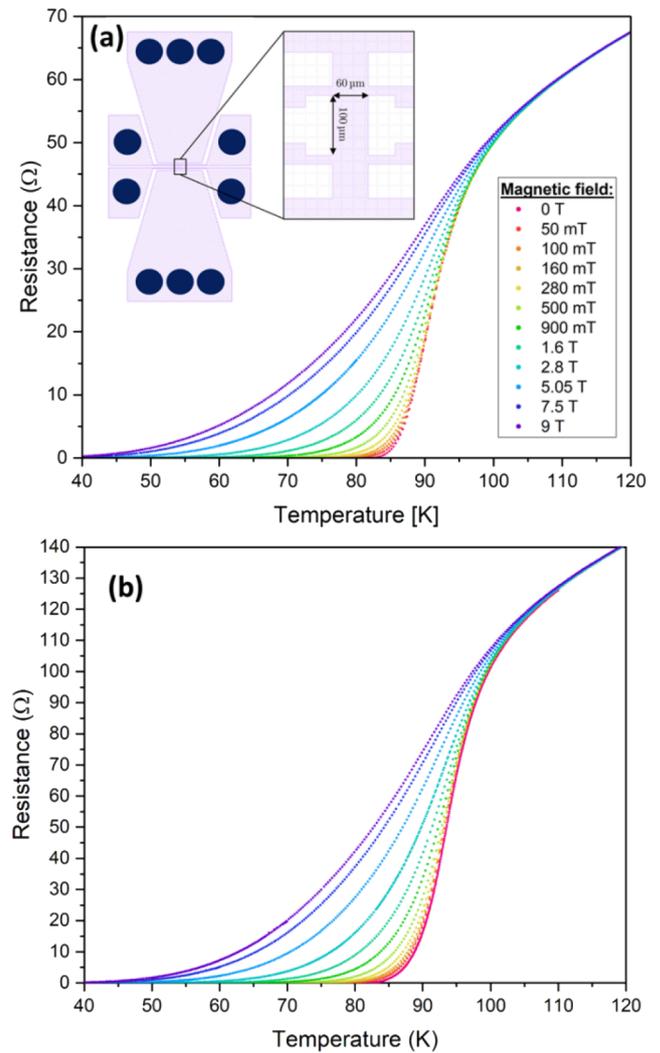

**Fig. 9.** Resistance versus temperature and magnetic field of (a) an optimized Bi-2212 film without impurities and (b) a Bi-2212 film with impurity structures. Films were grown on LAO substrates; film thickness was 45 nm. Field orientation was perpendicular to the sample surface. Lithographic pattern of films used for the four-point measurements ((a) inset).

The normalized phase transition temperature versus magnetic field $T_{c0}(B) / T_{c0}(B = 0)$ of the Bi-2212 films on LAO and of a YBCO film on STO is compared in Fig. 10. All samples were prepared by PLD and the data for the YBCO film were taken from ref. [33].

The Bi-2212 thin films reveal a rapid suppression of $T_{c0}$ with applied field $B$. The suppression is somewhat smaller for the phase-pure film compared to the film containing the impurity-phase. This behavior is in qualitative agreement with the higher critical current density $J_c$ and higher upper critical field $B_{c2}$ for the optimized, phase-pure Bi-2212 films. For comparison, the $T_c(B)$ reduction of the YBCO film is much slower. The different stability of the superconducting phase against external magnetic fields is connected to the much higher anisotropy of Bi-2212 as compared to YBCO and the segmented structure of vortices ("pancakes") in the Bi-based cuprate. Efficient pinning of vortices is therefore more demanding in the Bi-based material. The in-field critical current densities reported for YBCO are higher than that of Bi-2212, indicating the different flux pinning behavior of these HTS materials [36–39].

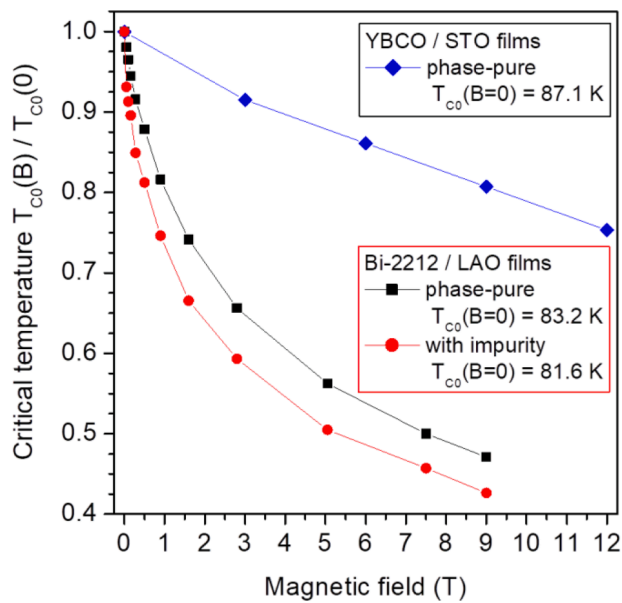

**Fig. 10.** Normalized critical temperature versus magnetic field of the Bi-2212 thin films on LAO and a YBCO thin film on STO (data for YBCO taken from ref. [33]). All films grown by PLD.

## 4. Conclusion

High-$T_c$ superconducting Bi-2212 thin films with thicknesses below 50 nm were deposited via PLD on different substrates (MgO, STO, LAO). The LAO substrates were found to be best suited for the preparation of very thin and smooth Bi-2212 films due to the good lattice match. Such c-axis oriented films were phase-pure, had a low surface roughness of 3 nm, and showed $T_{c0}$ up to 86 K and $J_c$(60 K) above 1 MA/cm$^2$. The drawbacks of this substrate choice are the structural phase transition of LAO at 550 °C creating defects and the observed decomposition of Bi-2212 into Bi$_2$Sr$_2$CaO$_x$ and CuO$_2$ far below the reported temperature of 820 °C. The "recycling" of used LAO crystal substrates has been studied and the influence of the recycling process on the deposited Bi-2212 thin films was determined. Chemical etching of the Bi-2212 layer and subsequent polishing of the crystal enabled to re-use the LAO substrates and to grow Bi-2212 thin films of high quality.

## CRediT authorship contribution statement

**Sandra Keppert:** Sample preparation (ceramics, thin film growth), sample characterization (XRD and resistivity measurements), Writing: Draft of paper. **Bernd Aichner:** Detailed measurements of superconducting thin film properties in magnetic field. **Rajdeep Adhikari:** Resistivity measurement of thin film at low temperature in magnetic field. **Bogdan Faina:** Resistivity measurement of thin film at low temperature in magnetic field. **Wolfgang Lang:** Conceptualization, Methodology. **Johannes D. Pedarnig:** Conceptualization, Methodology, Supervision, Writing – review & editing.

## Declaration of Competing Interest

The authors declare that they have no known competing financial interests or personal relationships that could have appeared to influence the work reported in this paper.

## Data availability

Data will be made available on request.

## Acknowledgments

We wish to thank R. Rössler for his previous work on optimizing the solid-state reaction technique for Bi-2212 ceramics. Furthermore, we acknowledge the outstanding support of our technical staff members H. Piglmayer-Brezina and A. Nimmervoll who are always willing to go above and beyond of what is expected of them. We thank A. Bonanni and J. Stangl at JKU Linz for providing measurement opportunities in their laboratories. This research was funded in whole, or in part, by the Austrian Science Fund (FWF) grant number I4865-N. For the purpose of open access, the authors have applied a CC BY public copyright license to any Author Accepted Manuscript version arising from this submission. R. Adhikari would like to acknowledge the project TAI-817 funded by FWF. This article is based upon work from COST Actions Super-QuMap CA21144 and Hi-SCALE CA19108 (European Cooperation in Science and Technology).